\documentclass[fleqn,twoside]{article}
\usepackage[headings]{espcrc2}
\usepackage[ruled,vlined]{algorithm2e}
\usepackage{tabularx,ragged2e,booktabs,caption}

\readRCS
$Id: espcrc2.tex,v 1.2 2004/02/24 11:22:11 spepping Exp $
\usepackage{graphicx, balance}
\usepackage[figuresright]{rotating}
\usepackage{graphicx}

\usepackage{verbatim}
\usepackage{amssymb}

\newcommand{\AmS}{{\protect\the\textfont2
  A\kern-.1667em\lower.5ex\hbox{M}\kern-.125emS}}

\title{\textbf{Distributed Painting by a Swarm of Robots with Unlimited Sensing Capabilities and Its Simulation}}

\author{Deepanwita Das\address{Department of Information Technology, National Institute of Technology, Durgapur 713 209 India, Contact: deepanwitadaptary@gmail.com \\},
Srabani Mukhopadhyaya\address{Department of Computer Science and Engineering, Birla Institute of Technology, Mesra, Kolkata Campus, Kolkata 700108 India, Contact: smukhopadhyaya@bitmesra.ac.in\\}}

\runtitle{Distributed Coverage by a Set of Swarm Robots}
\runauthor{Das D, et al.,}

\begin{document}
\begin{abstract}
This paper presents a distributed painting algorithm for painting a priori known 
rectangular region by swarm of autonomous mobile robots.
We assume that the region is obstacle free and of rectangular in shape. The basic approach is to divide the region into some 
cells, and to let each robot to paint one of these cells.
Assignment of different cells to the robots is done by ranking the robots according to their relative positions. 
In this algorithm, the robots follow the basic \emph{Wait}-\emph{Observe}-\emph{Compute}-\emph{Move} model together with the \emph{Asynchronous} timing model.
This paper also presents a simulation of the proposed algorithm. The simulation is performed using the Player/Stage Robotic Simulator on Ubuntu 10.04 (Lucid Lynx) platform.

\end{abstract}

\maketitle

\section{INTRODUCTION}

Distributed coverage of any polygonal region has been an important area of research over the past few years. Applications of covering a free space can be found
in the areas like automated humanitarian demining, lawn mowing and milling \cite{1}, sweeping \cite{2},
terrain mapping, space explorations, aerial reconnaissance, search and rescue of victims \cite{3} etc. Coverage of a particular region requires the robots to scan or pass over a designated region. 
When the robots cover or pass all the parts of that region, coverage is said to be complete. High quality coverage guarantees exhaustive 
coverage with minimum repetition. 
Each robot in a swarm, distributedly and simultaneously covering different parts of the area also minimizes time and cost of the work while 
increasing overall performance. \\
In this paper, one of such coverage problems is addressed. We consider a problem for painting a known 
rectangular region without any obstacle. The overall painting will be performed by a swarm of autonomous mobile robots. 
We assume that a set of $N$ swarm robots are initially deployed within the given rectangular region. 
The robots can be located at any place within that region.
These robots are assigned the responsibility to paint the whole region. Here, 
the proposed algorithm will be executed by each of the robots, to solve this problem collectively. 
We assume that the robots will work in a completely distributed environment. Painting a region is same as covering or scanning the region. From now on, the two words {\it coverage} and 
{\it painting} will be used interchangeably. \\
In this paper, the robots follow a basic model for computation which is known as \emph{wait}-\emph{observe}-\emph{compute}-\emph{move} model 
\cite{4} or CORDA model \cite{5}.  
The algorithms based on this basic \emph{wait}-\emph{observe}-\emph{compute}-\emph{move} model consists of a sequence of computational cycles. 
In every computational cycle, a robot executes the following four steps:\\
{\bf Wait}: A robot is initially in a waiting or idle state, but cannot stay infinitely idle.\\
{\bf Observe}: At any point of time a robot observes the positions of all other robots, asynchronously and independently from the other robots.\\
{\bf Compute}: Depending on the observations made in the previous step, the robot calculates its destination point 
based on its own position and the current locations of the other robots, etc.\\
{\bf Move}: The robot moves towards the computed destination.\\
The robots also use the {\it direction-only} model in which directions of both axes are common to all the robots, but the positive 
orientation of the axes are different. The robots follow {\it asynchronous} model in which they do not share any common clock and 
operate on independent computational computational cycles of variable length. The robots virtually divides the whole rectangular region into 
a number of non-overlapping cells (sub-region). Then each cell is assigned to a distinct robot which will be responsible for 
painting that particular cell. When each of the robots completes painting the cell assigned to it, the whole area will be painted or covered. Using 
the Player/Stage robotic simulator we have designed a controller program that simulates the proposed algorithm. Each robot will execute the controller program 
repeatedly until all the robots completes their assigned job. \\
Most of the previous works consider the presence of obstacles within the area to be covered.
Although, all these algorithms are also applicable for the areas without obstacle, using these algorithms 
for a region without obstacle makes the solution unnecessarily complicated.  
Our proposed algorithm shows that the coverage problem without any obstacle can be solved in a more realistic way
using robots of simple nature. We assume that there is neither any central authority nor any external control over 
the robots. Moreover, during execution of the proposed algorithm robots do not need to communicate among themselves. 
Each robot assumes a local co-ordinate system and all the computations carried out by the robots are according 
to their respective local co-ordinate system. Further, to provide a more realistic solution 
to the painting problem, we assume that a robot can be in two different states, {\it active} state and 
{\it sleep} state. However, from a sleep state a robot becomes active within a finite amount of time. \\
In Section~\ref{liturature}, we discuss the related research work done in this area. Section~\ref{modelAssump}
introduces the problem definitions, models and assumptions for the solution of the problem. Section~\ref{covAlgo}  is having two subsections. In the first
sub-section, the painting algorithm is discussed and in the second sub-section the correctness of the algorithm is established.
The simulation of the proposed algorithm is discussed in Section~\ref{Simulation} and we conclude our study in Section~\ref{Conclusion}.

\section{RELATED WORK}\label{liturature}

Over last few years a large amount of research work has been reported on Multi-Robot Coverage problem \cite{3}, \cite{6}, 
\cite{7}, \cite{8}, \cite{9}, \cite{10}, \cite{11}, \cite{12}. Different approaches are followed to cover a given region with or without obstacles within it. 
Most of the related works consider the Boustrophedon Decomposition \cite{13} approach, which divides the target space into sub-regions called 
cells where each cell can be covered by the robots with simple back-and-forth motions. Different approaches are followed to 
define these cells. For example, Canny et al., \cite{14} defined the cells by sweeping a slice
(a one dimensional line) through the configuration space. \\
Latimer et al., \cite{15} introduced a multi-robot coverage algorithm based on the same planer cell-based approach for a
single robot. Robots move in a team, they communicate their state and share information. The robots in a team move by maintaining an horizontal formation, like a rake. 
The central issue of this paper is when to divide and merge teams so that all the cells in the region are covered. 
This is solved by critical point detection method. Whenever the team detects any obstacle in the slice, 
it is unable to continue as one unit. Thus, a critical point is detected at that point and the cell is divided into sub-cells. 
Depending on the types of critical points, the team is also divided into sub-teams to cover the sub-cells. 
After covering the sub-cells adjacent to the obstacle the sub-teams rejoins. 
In a team based approach communication, coordination and synchronization are required 
among the team members, which make the robot more complex. 
However, in this approach repeated coverage may occur \cite{15}. \\
Rekletis et al., \cite{6}, \cite{10}, used same planar cell-based decomposition and also provided extensions to handle how 
teams of robots cover a single cell and how are they re-allocated among different cells. 
Considering \emph{communication} among robots as a key issue, two algorithms were proposed. (i) \emph{Team-based coverage} 
for restricted communication and (ii) \emph{Distributed coverage} for unrestricted communication. In \cite{6}, the communication among the robots is restricted to their line-of-sight. In this paper,
the robots in a team are categorized as explorers and coverers. The basic idea is that, when two explorers (while moving 
along upper and lower boundaries of the region) loose their line-of-sight due to the presence of an obstacle determines the 
position of the critical point. At this point two sub-cells, adjacent to the obstacle, are generated which are then 
covered by the two sub-teams. Once again two robots in each of these sub-teams will be categorized as explorers. 
Such form of coverage requires the coverers to move in team formations, which may be accomplished in a 
variety of ways \cite{16}. The coverers may cover some region previously covered by the explorers resulting repeated coverage. 
Moreover, maintaining the line-of-sight communication among the explorers or coverers introduces considerable amount of complexity.\\
In unrestricted communication, the area to be covered is divided into a number of virtual strips equal to the number of robots. 
The robots are deployed at regular intervals along one side of the region to be covered and they 
start exploration of strip 
boundaries using the cycle algorithm developed in single-robot Morse Decomposition \cite{17}. 
Same cycle algorithm is used for coverage also. During exploration, the robot gathers the knowledge of 
critical points and steiner points (points that represent strip boundaries). Depending on that, the robot builds 
a {\it global reeb graph}, which is shared and updated by all the robots. After completion of exploration, robots 
immediately start coverage of known strips and update the {\it reeb graph} accordingly. If a robot is unable to reach 
any space within its strip, it re-allocates the unreachable part to other robots by calling an auction mechanism. 
It selects the robot which can explore the unreachable part at a lower estimated cost. This algorithm is efficient 
but complex. Moreover, the deployment mechanism of the robots is not realistic.\\
Kong et al., \cite{18} uses the Boustrophedon Decomposition approach. The robots are initially distributed inside a region and each one is allocated 
a virtually bounded cell of that region. The region is divided into several fixed width cells. The robots  
determine whether the cell being covered is divided into disconnected parts due to the obstacles. 
Each robot uses an {\it adjacency graph} that represents its current cell and the adjacent cells to be covered. 
The graph also represents the disconnected parts of a cell due to the presence of any obstacle. During execution of the
algorithm, new cells are added to the {\it adjacency graph}. When a robot completes covering a cell, it shares 
the new graph information among all the robots. A robot can communicate graph information to all other robots without any restriction.
Upon receiving the graph information other robots update their own graph. A robot then selects the next uncovered cell in its {\it adjacency graph}. 
The algorithm is good for coverage but maintaining and sharing information of all covered/uncovered cells among all the robots requires 
large amount of memory, making the robots no more oblivious.  \\
Most of the previous works consider the presence of obstacles within the area to be covered. None of the above mentioned 
algorithms is based on CORDA model \cite{5}. All the robots are assumed to be {\it synchronous} and {\it active} throughout 
the process. The robots agree on the direction and orientation of the axes. Deployment of 
robots within the region is also not random. Communication among robots is absolutely necessary in all these 
approaches. Repeated coverage are likely to occur due to loss of the communication among the robots. In the team based approaches, communication, 
coordination and synchronization among the team members involves great complexity. In \cite{10}, \cite{18}, maintaining the 
graphs requires large amount of memory and some times a centralized control on robots is also required. \\ 
Our proposed algorithm shows that the coverage problem without any obstacle can be solved in a more realistic way
using simple robots instead of the complicated one. The robots actually work in a totally distributed way, independently from other robots. They are distributed randomly within 
the region to be covered. The robots do not communicate among themselves. Each robot does all the computations based on the information gathered with 
respect to its local co-ordinate system. We assume that the local co-ordinate systems agree on the direction but orientation. We also consider the CORDA model, 
a standard computational model for 
distributed robot system. Our algorithm is based on \emph{asynchronous} model, where robots may not be active always and operates on 
independent computational cycles of variable lengths. We assume that a robot can be in two different states, {\it active} state and 
{\it sleep} state. However, from a sleep state a robot becomes active within a finite amount of time. This guarantees the finite time completion of the painting task. 

\section{MODEL, ASSUMPTIONS AND PROBLEM DEFINITION}\label{modelAssump}

Before going to describe the algorithm, let us discuss the assumptions and the models used. We also introduce the terminologies used in this paper. \\
Our problem is to paint a given rectangular region by a swarm of $N$ robots. These robots are initially deployed randomly within the rectangular region, 
which is to be painted. Robots may occupy any position within the region. We assume that no two robots occupy the same position. The robots we 
consider here are relatively weak, simple and assumed to have the following characteristics \cite{19}:
\begin{enumerate}
\item \emph{Identical and Homogeneous} - All the robots are identical in all respect, specially, they have the same computational capability. 
All the robots are assumed to be point robots with unlimited visibility. However, we assume that each of them is having a sensing zone of radius 
$\eta$ ($\eta$ is small). That is, if a robot is required to carry out some job related to a particular position (collection of information about that 
position, or painting that position etc.), instead of actually reaching the position, the robot can also carry out the job from a distance of $\eta$ away from it. 
In case of painting, as if, each of the robot is carrying a paint brush of length $\eta$. So, while painting, if a robot moves in a straight line, 
a rectangular strip of width $2\eta$ about that line will be painted as shown in the Fig. \ref{fig:0}.    

\item \emph{Autonomous} - There is neither any central authority nor any external control over the robots. Thus the robots work in  
completely distributed manner, asynchronously and independently from other robots. They do not even communicate among themselves.
\begin{figure}[h]
\centering
\includegraphics[height=20mm, width=45mm]{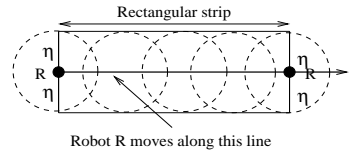}
\caption{The area sensed by a robot while moving along a straight line}
\label{fig:0}
\end{figure}
\item \emph{Mobile} - All robots are allowed to move on a plane.

\item \emph{Computation Model} -  Here we follow the basic \emph{Observe}-\emph{Compute}-\emph{Move} \cite{4} model. 
A computational \emph{cycle} is defined to be  a sequence of  {\it observe}, {\it compute} and  {\it move} steps. Each of the robots executes same 
instructions in all the computational cycles. Once a robot completes one computational cycle, it starts executing the next one. The actions taken 
by a robot in  {\it compute} and {\it move} steps, entirely depend on the observations made in  {\it observe} step. In some situations, 
an observation might lead a robot not to change its position in {\it move} step. In such cases the robot seems to be idle, though it is actually 
executing all the three steps.

\item \emph{Oblivious} or \emph{Memoryless} - Robots do not retain any information gathered in the previous computational cycle. 
In every computational cycle, a robot starts computing from very beginning depending only on the positions of the other robots 
observed at that computational cycle. 
\end{enumerate}

The robots can have two states:  \emph{active} and \emph{sleep}. In \emph{Active} state, the robots are alive and executing continuously the 
computational cycles. In \emph{Sleep} state, robot is not active and doing nothing. This state is like {\it power off} state. It is assumed that 
a robot cannot \emph{sleep} infinitely and it would become active within a finite amount of time. We also assume that change of state of a robot 
takes place independent of the other robots.\\
The {\it painting} operation considered here is assumed to be an \emph{Atomic} operation. Once a robot starts painting the assigned cell, 
it completes its job without any further interruption. During painting, a robot cannot switch over to the {\it sleep} state also.\\

The models considered here are as follows:
 
\textbf{Asynchronous model:} Robots operate on independent cycles of variable lengths. They do not share any common clock \cite{19}.\\
\begin{figure}
\centering
\includegraphics[height=35mm, width=80mm]{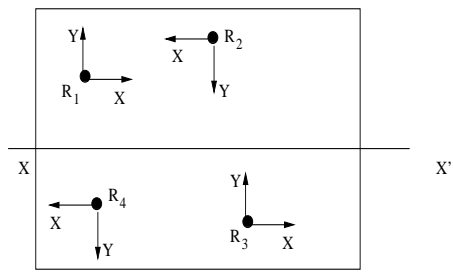}\\
\caption{Positive orientation of the local axes of $R_1$ and $R_3$ are just the reverse of that of the robots $R_2$ and $R_4$}
\label{fig:1} 
\end{figure}
\textbf{Direction only:} Directions of both axes are common to all the robots, but the positive orientation of the axes may be different \cite{19}. 
Here, we assume that x-axes of the robots are parallel to the  known common reference line. Therefore, the direction of x-axis is common to all the 
robots but the robots may have different views of the positive orientation of the axis.  However, it is assumed that the direction of the positive 
y-axis is $90^{\circ}$ counterclockwise to the positive direction of the x-axis. Thus, direction of y-axis is also common to all the robots, except 
possibly the positive orientation. Each robot has its local co-ordinate system. All the robots would assume that they occupy the position $(0,0)$ 
with respect to their local co-ordinate system. Further, we assume that these various co-ordinate systems might not share a common scale. 
Fig. \ref{fig:1} shows the local co-ordinate systems of four robots $R_1$, $R_2$, $R_3$, and $R_4$, and the common reference line $\overrightarrow{XX'}$.

\section{ALGORITHM}\label{covAlgo}
The first part of this section describes the proposed algorithm. The correctness of the Algorithm is established in the second part.

\subsection{Algorithm for Painting} \label{Algo}
It is assumed that the region to be painted is a rectangular region and no obstacles are there within the region. Further, we assume that all 
the $N$ robots are enclosed within the region. Without loss of generality, we assume that the common reference line be parallel to one side of 
the rectangular region.\\

The algorithm completes the painting job in two phases. In {\it Phase I}, a robot calculates the strip to be painted by it and moves to the starting position 
where from it can start painting. 
In this phase, as soon as a robot $R$ becomes alive it performs the following computational cycle {\it observe-compute-move}.
After completing one such computational cycle, the robot $R$ would again start another cycle and continue in this way 
until it reaches to the starting position for painting or it goes to {\it sleep} state again. After completion of {\it Phase-I}, a robot starts {\it Phase-II} 
in which it actually paints the assigned strip.\\

\noindent {\bf Algorithm {\it Paint}}\\

The following steps are executed by the robot $R$.\\
\begin{algorithm}
{\bf Phase-I:}
\hspace*{0.1 in}{\bf do} \\
\hspace*{1.0 in} Observe\\
\hspace*{1.0 in} Compute \\
\hspace*{1.0 in} Move\\
\hspace*{0.7 in}{\bf while}(the robot $R$ is alive and the   
\hspace*{0.7 in} robot has not yet reached the 
\hspace*{0.7 in} starting position for painting) \\

{\bf Phase-II:} The robot paints its assigned strip.\\
\end{algorithm}

{\bf Phase-I:} In this phase the robot calculates its strip and moves to the starting position for painting. Let us discuss the steps in details:\\

{\bf Observe}\\
According to the local co-ordinate system, the robot  $R$ first observes the positions of all other robots. Let the co-ordinates of other robots be 
$(a_1, b_1)$, $(a_2, b_2)$,  $\cdots$, $(a_{N-1}, b_{N-1})$, whereas, its own co-ordinate be $(0, 0)$.  It is to be noted here that some of 
these $a_i$, $b_i$ values might be negative also.\\

{\bf Compute}\\
\underline{Step 1}:\\
According to the values of $y$-co-ordinates, the robot $R$ orders all the robots (including itself) so that the robot having the largest value 
of $y$ co-ordinate will have the highest rank, that is, $N$. Without loss of generality, let us assume that after 
sorting the co-ordinates of $N$ robots 
be $(x_1, y_1)$, $(x_2, y_2)$,  $\cdots$, $(x_{N}, y_{N})$, so that $y_1$ $\leq$ $y_2$ $\leq$ $y_3$ $\leq$ $\cdots$ $\leq$ $y_N$. The robot having 
the co-ordinate $(x_i, y_i)$ have the rank $i$ and from now on the robot will be mentioned as $R_i$, $1 \; \leq i \;\leq N$.
In case of a tie, the values of $x$ -co-ordinate of the robots are to be considered. The robot having lower $x$-co-ordinate would have the lower rank. 
As we have assumed that no two robots can occupy the same position, two robots having the same $y$-co-ordinate cannot have identical $x$-co-ordinate.\\
In this way, the robot $R$ would determine its own rank. Let the rank of $R$ be $k$. From now on $R$ and $R_k$ will be used interchangeably.\\

\underline{Step 2:}\\
According to the local co-ordinate system of $R$, let the upper boundary of the region to be painted be at a vertical distance  $s$ and the 
lower boundary be at $f$. Since all the robots are enclosed within the area, $s \; \geq \; 0$ and $f \; \leq \; 0$.  The whole rectangular area 
will be divided into $N$ equal horizontal strips of height $({s-f \over N})$. The top most  (according to the local co-ordinate system of $R$) 
strip will be considered as the $N^{\rm th}$ strip and the bottom most one will be considered as the first strip. Now the robot $R_k$ will identify 
the $k^{\rm th}$ strip by computing its upper and lower boundary as  $f + (k-1) * ({s-f \over N})$ and $f +  k * ({s-f \over N})$. The $k^{\rm th}$ 
strip will be colored by $R_k$. Each robot would start the coloring from the bottom left corner of the assigned strip. Accordingly the robot would 
compute its destination.\\
On the way towards their destination, robots maintain their relative ranking. It means, while moving, robots should not cross vertically 
any other robot even if their routes do not intersect each other. In other words, to reach the destination, if a robot is going to gain a vertical 
height higher(lower) than a robot of higher(lower) rank (that is, it is crossing another robot which would affect the relative ranking), it would 
stop at an $\epsilon$ (pre-defined small quantity) distance from that height and  would wait for that other robot to move on. \\
\begin{figure}[h]
\centering
\includegraphics[height=30mm, width=70mm]{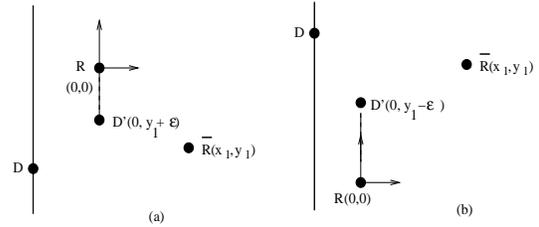}
\caption{Secondary destinations $D'$ w.r.t robot R, where $D$ is the actual destination}
\label{fig:com}
\end{figure}
Due to the rule stated above, the robot $R$ may need to take a halt before reaching its final destination, the bottom-left corner of the 
assigned strip. In this {\it compute} step, the robot $R$ should verify this situation and if required, it would recalculate the position of the halt. 
We call this as the {\it secondary} destination. Suppose, to reach the final destination $D$, $R$ has to vertically cross another robot $\overline{R}$ 
which is at a point ($x_1$, $y_1$) as shown in Fig. \ref{fig:com}. According to the given rule, $R$ would stop at a vertical height of $y_1\pm\;\epsilon$. Therefore, the modified 
destination of $R$ would be ($0$, $y_1\pm \epsilon$). Fig. \ref{fig:com} shows two possible cases. In Fig. \ref{fig:com}(a), to reach the 
destination $D$, the robot $R$ has to move in the vertically downward direction and then it requires to cross $\overline{R}$. Therefore, its 
secondary destination would be ($0$, $y_1+\epsilon$). In Fig. \ref{fig:com}(b), to reach the destination $D$, the robot $R$ has to move in 
the vertically upward direction and then it requires to cross $\overline{R}$. Therefore, its secondary destination would be ($0$, $y_1-\epsilon$). \\
This {\it compute} step is terminated as soon as the robot computes its destination, final or secondary.\\ 

{\bf Move}\\
After identifying the assigned strip, the robot would start moving towards its destination point, the bottom left corner point of the assigned strip. 
Actually the robots do not need to reach the exact height of its destination due to its sensing ability. It is sufficient to reach a height, which is 
at a distance $\eta$ (above/below) away from the final destination (as discussed in Section \ref{modelAssump}).\\
It is to be noted here that, though the final destination of a robot is the bottom left corner point of the assigned strip, sometimes, to preserve the 
relative ranking, robots may need to wait at certain height for some other robot to move on.\\
A robot would always move in vertical direction first, after acquiring the vertical height of the final destination, the robot would then move along 
horizontal direction to reach the final destination. Thus, to reach the secondary destination, a robot moves only in vertical direction.\\
Depending on whether a robot reaches its final or secondary destination, the following two courses of actions would be taken by the robot:  \\

(i) As soon as a robot reaches the secondary destination, this {\it move} state terminates. That is, the current computational cycle will be terminated 
and the robot will again start a new computational cycle with {\it observe} state.  \\

(ii) Once the robot reaches its final destination before starting the painting in {\it Phase-II}, it would check whether there is any other robot present in 
its assigned strip or not. There may be two possible cases: \\

{\it Case I : The robot finds another robot in its own strip}\\
If the robot $R$ finds another robot, say $\overline{R}$,  present in its assigned strip, 
it would wait for that other robot to move on. It will keep on executing the sequence of {\it observe-compute-move} steps until the strip become empty. In this situation,
there will not be any movement of the robot as it has already reached its {\it final} destination. \\

{\it Case II : The strip is empty}\\
If there is no other robot in the strip and the strip is empty, the robot would go to {\it Phase-II} for painting. \\

At any point of time, if the robot $R$ finds another robot $\overline{R}$ at the same vertical height (which might occur at the starting time, if 
initially they are at the same height), then depending on the rank of $\overline{R}$ and that of itself, $R$ decides its next course of action as follows: \\

{\bf Case A :} The rank of $R$ is greater than that of $\overline{R}$ and the destination of $R$ is in the positive direction, w.r.t. its local co-ordinate system.\\ \\

{\bf Case B :} The rank of $R$ is less than that of $\overline{R}$ and the destination of $R$ is in the negative direction, w.r.t. its local co-ordinate system. \\

For both the cases A and B, $R$ would break the tie and would move first towards its destination point.\\

{\bf Case C :} The rank of $R$ is greater than that of $\overline{R}$ and the destination of $R$ is in the negative direction, w.r.t. its local co-ordinate system. \\

{\bf Case D :} The rank of $R$ is less than that of $\overline{R}$ and the destination of $R$ is in the positive direction, w.r.t. its local co-ordinate system. \\

For both the cases C and D, $R$ will wait for $\overline{R}$ to move first towards its destination point. \\

{\bf Phase-II:}  In this phase the robot start painting the assigned area. As painting is considered as an \emph{Atomic} operation, 
the robot would complete the job successfully without any interruption and at the end, it would generate a signal that its job is done. 

\subsection{Correctness of the Algorithm}\label{correctness}
\textbf{Observation 1:} {\it Throughout the process, relative ranking of the robots computed by several robots  are same upto a reversal of order. 
In other words, if the robots $R_1$ and $R_2$ compute the rank of a robot $R$ as $i$ and $j$ respectively, then either $i=j$ or $i= N+1-j$ and this would 
remain same throughout the algorithm.}\\
\begin{figure}[h]
\centering
\includegraphics[height=60mm, width=65mm]{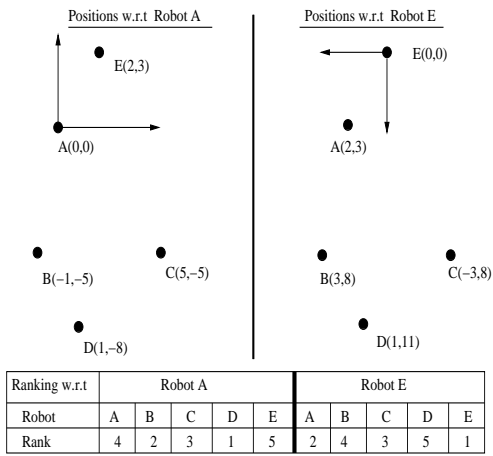}
\caption{Relative rankings w.r.t. robots A and E are same upto a reversal of order}
\label{fig:2}
\end{figure}\\

\emph{Proof}: If the orientations of the local axes of $R_1$ and $R_2$ are identical then the ranking of the robots would be same. Otherwise, 
if the orientations are reverse, then the relative ordering would be same but in reverse order. Thus $i =  N+1-j$. Fig. \ref{fig:2} shows that the 
relative ranking of the robots with respect to two different robots having opposite orientation of their axes, are just the reverse. In this figure, 
an example of five robots are shown. Here, robots A and E are having opposite orientations of their axes. Positions of all the robots and the relative 
ranking of all the robots with respect to robots A and E are shown in the figure.\\
A robot computes the ranks of all other robots w.r.t. their vertical distances from its local $x$-axis. So the relative ranking of the robots would 
remain same throughout the algorithm as the vertical movement of the robots is so restricted that none of the robots would vertically cross any other 
robot. If two robots are starting from the same vertical height, their relative ranking will be determined by their $x$-coordinates. In case of such a tie, 
the robots start moving towards their destination following the rule given in {\it move} step, which retains their relative ranking. Once a robot starts 
moving, this tie will be broken and this situation will never occur again.\\

\noindent\textbf{Observation 2:} {\it The assignments of cells (for painting) to the robots as computed by different robots are same and it would remain 
same throughout the whole process.}\\

\emph{Proof}: It is obvious when the orientations of the local axes of the robots are same. Let $R_1$ and $R_2$ be two robots whose axes are oriented 
just in opposite direction. Let $R_1$ computes the rank of a robot $R$ as $k$ and assigns the $k^{\rm th}$ cell to $R$. Now, $R_2$ would compute the rank 
of $R$ as $N+1-k$ and it would assign the same cell to $R$ which according to local co-ordinate system of $R_2$ is considered to be $N+1-k$.  The position of 
the cells are fixed and the relative ranking of the robots remain same throughout the algorithm. Hence, assignment of strips to the robots remains invariant 
with respect to any computational cycle.\\

\noindent\textbf{Observation 3:} {\it The movements of robots are collision free.}\\ 

\emph{Proof}: Throughout the algorithm, two robots can never be at the same vertical height at the same time, except possibly, at the starting time. 
If initially the robots are at the same height, the tie will be broken by the rules given in {\it move} step. 
Once the tie is broken, they will never be at the same height again, during their vertical movement.\\ 
\begin{figure}[!h]
\centering
\includegraphics[height=60mm, width=65mm]{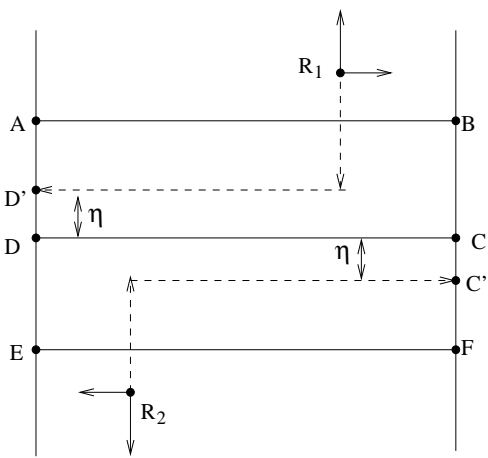}
\caption{Deadlock avoidance using sensing ability}
\label{fig:12}
\end{figure}
After computing the destination, each robot would first move vertically to reach the height of the final destination. Once they reach that height, 
they start moving horizontally. Thus, if the destinations of two robots are at different heights, the question of collision during their horizontal 
movements does not arise at all.\\
\begin{figure}[!h]
\centering
\includegraphics[height=45mm, width=70mm]{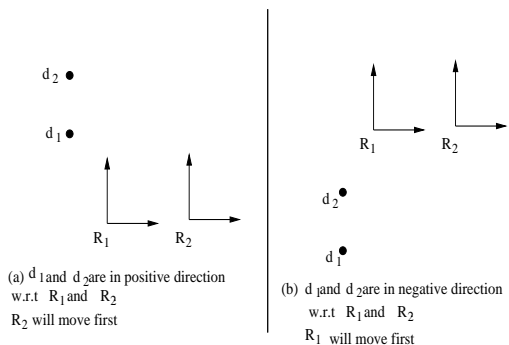}
\caption{Robots have same orientation}
\label{fig:4}
\end{figure}
There is one special case where the final destinations may have same vertical heights for two different robots. As shown in Fig. \ref{fig:12}, 
two robots $R_1$ and $R_2$ are assigned to paint two consecutive strips, $ABCD$ and $CDEF$ respectively. Let us assume $R_1$ and $R_2$ have opposite 
orientation of their local axes as shown in the figure. The bottom-left corner point of the strip $ABCD$, w.r.t. $R_1$ is the point $D$. Whereas, 
the bottom left corner point of the strip $CDEF$ is at $C$ according to $R_2$. This shows that the final destination of both the robots $R_1$ and $R_2$ are 
at the same vertical height. Therefore, their horizontal movement will be along the same line $CD$, which may cause collision. 
However, due to their sensing ability (as discussed in Section \ref{modelAssump}), $R_1$ would start its job from the point $D'$ and $R_2$ from $C'$, which are at a $\eta$ distance away from the line $CD$. Thus, here also, the two robots will not move along the same horizontal 
line to reach their destination for starting the job.\\ 

There will not be any collision in {\it Phase-II} also. Before starting the actual painting in {\it Phase-II}, a robot verify whether the strip is empty or not. If it finds 
another robot in that strip it will wait till it becomes empty. Once the strip becomes empty there is no possibility of any other robot to enter the strip again.
This is due to the fact that none of the robots vertically cross any other robot.\\

\textbf{Observation 4:} {\it The four rules stated in the {\it move} step lead to take the robots a non-conflicting decision regarding tie-breaking.}\\   

If at the initial situation, two robots are at the same height (but definitely in two different positions), the robot having the higher rank would start 
moving first, if their destinations are in the positive direction. If their destinations are in the negative direction, then the robot having the lower 
rank would start moving first. If their destinations are in opposite direction, then there wouldn't be any restriction in vertical movement.\\                        
Consider Fig. \ref{fig:4} where both $R_1$ and $R_2$ having same orientation. $d_1$ and $d_2$ are the destinations of robots $R_1$ and $R_2$ respectively. 
According to both the robots the rank of $R_1$ is less than the rank of robot $R_2$. In Fig. \ref{fig:4}(a) both of their destinations are in the positive 
direction then as per rule, the higher ranked robot $R_2$ will move first. In Fig. \ref{fig:4}(b) both of their destinations are in the negative direction. 
As per rule, the lower ranked robot $R_1$ will move first.\\
\begin{figure}[!h]
\centering
\includegraphics[height=40mm, width=70mm]{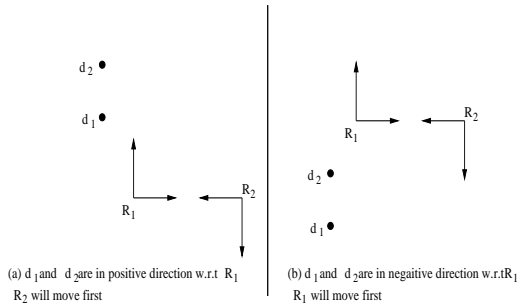}
\caption{Robots have opposite orientation}
\label{fig:5}
\end{figure}\\
Consider Fig. \ref{fig:5} where both $R_1$ and $R_2$ having opposite orientation. According to the local coordinate system of $R_1$ and $R_2$, 
both will rank itself as lower.  Due to opposite orientation, if the destinations are in positive direction according to $R_1$ then it is in negative 
direction according to $R_2$ and vise versa. In Fig. \ref{fig:5}(a), both the destinations are in positive direction w.r.t $R_1$. So, according to 
$R_1$, the higher ranked robot $R_2$ will move first. But according to $R_2$ the destinations are in negative direction, so as per rule the lower ranked 
robot $R_2$ will move first. This shows that the same decision will be taken by $R_1$ and $R_2$. \\
\begin{figure}[!h]
\centering
\includegraphics[height=30mm, width=65mm]{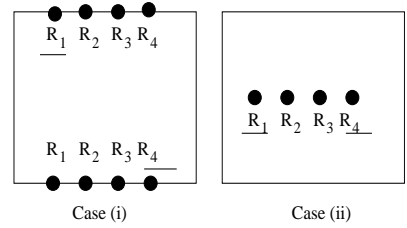}
\caption{Underlined robots are tie-breaking robots}
\label{fig:6}
\end{figure}
Similarly, in Fig. \ref{fig:5}(b), both the destinations are in negative direction w.r.t $R_1$. So, according to $R_1$, the lower ranked robot 
$R_1$ will move first. But now according to $R_2$ the destinations are in positive direction, so as per rule higher ranked robot will move first 
which is $R_1$ according to the local coordinate system of $R_2$. So, in both the cases same robot will move, and the tie will be broken without any conflict.\\ 

\noindent\textbf{Observation 5:} {\it The process would start within finite amount of time.}\\   
 
\emph{Proof}: According to our assumption, a robot cannot be in {\it sleep} state for an infinite amount of time. Once they become {\it alive}, 
they start moving provided their movement would not affect the relative ranking. Only in case of a tie, i.e., if initially the robots are at the same height, 
there will be an inter-dependency among these robots. If a particular robot does not move first all other have to wait for it and so on. The robot having the 
highest rank and the lowest rank usually does not have any restriction on their movement and thus as soon as they become live the process would start. 
We can think of an extreme situation when all the robots are at the same height. We can subdivide this situation into following two cases:\\

{\bf Case-1:} All the robots are at a same height and they are along a boundary of the region. In this case, if a robot identify itself (according to its local 
co-ordinate system) (1) at the lower boundary of the region and having the highest rank, or (2) at the upper boundary of the region and having the lowest rank, 
then the robot will not have any restriction on its movement and it would break the initial barrier. We call these robots as \emph{tie-breaking} robots. 
If the robots are on the upper boundary, the left-most one and if they are on the lower boundary, the right-most one will be the \emph{tie-breaking} robot.\\

{\bf Case-2:} All the robots are at a same height from the common reference line but they are not along any boundary of the region. Here, 
both the robots having lowest rank and highest rank will not have any restriction on their movement and they would break the initial barrier. 
Fig. \ref{fig:6} shows both the cases where the tie-breaking robots either having highest or lowest rank.\\
Once a robot break the initial barrier, all other robots start moving in turn. Thus, within finite amount of time the process would start. \\

\textbf{Result} \emph{The painting will be completed within a finite amount of time}.\\

\emph{Proof}: Combining all the above observations, and the fact that a robot cannot be in sleep state for an infinite amount of time and the 
painting operation is an atomic operation, we can conclude that painting will be completed successfully within finite amount of time.

\section{SIMULATION AND EXPERIMENTAL RESULTS}\label{Simulation}

\subsection{Simulation}
The simulation has been conducted based on variety of co-ordinates and orientations of the robots using the Player/Stage multi-robot simulation software. 
The Player (Version 3.2.2) and Stage (version 3.0.2) softwares have been configured on Ubuntu 10.04 (Lucid Lynx) platform with support of Intel Core2Duo Processor 
with $3.00$ GHz speed and 2.00 GB RAM. \\
Some robots are deployed randomly inside a priori known bounded rectangular region. 
The initial location  of these robots are generated randomly with respect to the
global co-ordinate system keeping in mind that all of them should be located inside the rectangular region. 
During the execution of the algorithm, a robot acts with respect to its own local co-ordinate system.
The dimension of the rectangular area is fixed and known. Moreover, the orientation of the robots are also randomly selected. 
The orientation of any robot may be either positive or negative (w.r.t. the global co-ordinate system) and is represented by 
$P$ or $N$ accordingly. In the tables given below, a robot is represented as ($X$, $Y$, $O$), where $(X, Y)$ is its coordinate and $O$ is 
its orientation. In the simulation, all robots are assumed to start execution of the algorithm at the same time. \\

\begin{figure}[!h]
\centering
\includegraphics[height=60mm, width=70mm]{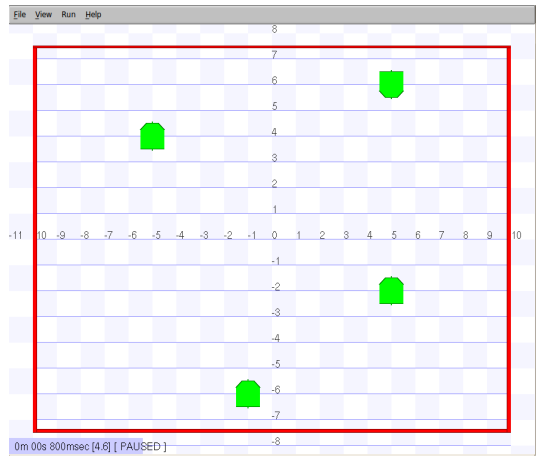}
\caption{The world with four robots with co-ordinates and orientation as (5,6,N), (5,-2,P), (-5,4,P) and (-1,-6,P)}
\label{fig:3}
\end{figure}

\begin{figure}[!h]
\centering
\includegraphics[height=60mm, width=70mm]{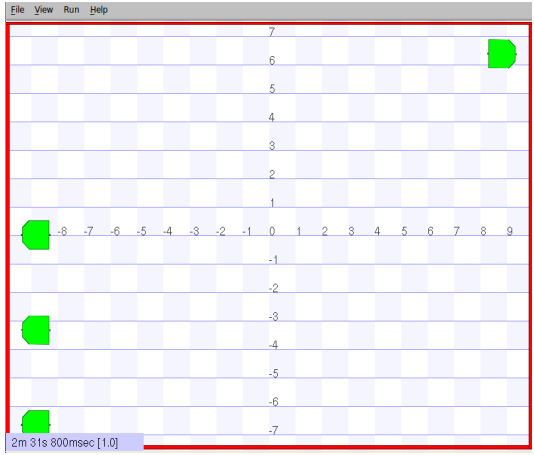}
\caption{Final destinations reached by the robots to start painting}
\label{fig:16}
\end{figure}

At first, a robot detects the boundary walls and all other robots within the area.
It calculates its rank based on the positions of the other robots. Next,  the robot calculates its assigned cell to be painted and 
the corresponding starting location. Finally it moves to that starting location, where it can start painting. 
Fig. \ref{fig:16} shows the robots (initial positions of these robots are shown in Fig. \ref{fig:3}) at their final locations from where they
start painting. The actual painting operation is not simulated as 
it is implied that once a robot reaches its final destination, it would be able to complete the painting within finite amount of time by following 
simple back and forth motion. \\
The total time of completion of the whole job is calculated as the sum of the time taken by the last robot to 
reach its final destination and the actual painting time required by the robots to paint the respective strips. 
Thus, $Total \; Time(T)$ = $t_1$ + $t_2$, where $t_1$ represents the time taken to complete {\it Phase-I} and $t_2$ is that of {\it Phase-II}.
$t_2$ can be estimated as ${{LB}\over {Nv}}$, where $L$ and $B$
are the length and breadth of the rectangular area respectively, $N$ is the total number of robots and $v$ represents the velocity of the 
robots.\\

\subsection{Simulation Environment}
The robots used in this simulation are all green in color and having $(1 unit \;\times\; 1 unit \;\times\; 1 unit)$ dimension. The robots are equipped with the following devices:
\begin{itemize}
 \item Infrared Laser sensors: Attached at the left and right side of the robot with sensing range upto $40$ units, $180$ scan lines and $180^{\circ}$ field of view and $180$ samples.
 \item The Blobfinders: Attached at the left and right side of the robot and capable to recognize two colors red and green and image of size 160 x 120 square unit. It is having a sensing 
range of 40 unit and $180^{\circ}$ field of view.
\end{itemize}

The world or environment of the simulation is considered as rectangular in shape with length as $30$ units and breadth as $40$ units. The whole rectangular area 
is bounded by a red colored boundary with the width of $1$ unit.\\
Each robot executes a controller program which implements the proposed algorithm. The controller program is written in C++ programming language. 
It uses the $libplayerc++$ library to communicate with the Stage. The controller program sequentially performs the {\it observe-compute-move} steps. 
In the controller program, each of the robot is programmed as a thread. Each thread independently and individually executes the controller program. 

\subsection{Results}

We have performed different tests by varying the (i) total number of robots, (ii) initial positions of the robots and 
(iii) orientations of the robots. In all the tests, the controller program was successfully completed within a 
finite amount of time. All the robots successfully reached their final destinations without any collision. 
\begin{table}[ht]
\captionof{table}{Total number of robot = 4}\label{table:1} 
\begin{center}                  
\small\addtolength{\tabcolsep}{-5pt}                         
\begin{tabular}{|c|c|c|c|c|}             
\hline                      
Robot & Instance & Instance & Instance & Instance \\  [0.5ex]   
      & 1 & 2 & 3 & 4 \\  [0.5ex]    
      & (X,Y,O) & (X,Y,O) & (X,Y,O) & (X,Y,O) \\
\hline                            
1 &(6,-3,P) &(13,4,N) &(10,3,N) &(12,9,N)\\               
\hline  
2 &(5,4,P) &(5,4,N) &(-7,2,N) &(12,-8,P)\\
\hline   
3 &(-5,11,P)&(-7,8,N)&(12,8,P)&(-14,9,P)\\ 
\hline  
4 &(-3,-5,P)&(-1,-5,N)&(-9,-5,P)&(-14,-8,P)\\
\hline
$t_1$ (min:sec) & 2:42 & 2:44 & 2:40 & 2:36\\          
\hline  
Avg($t_1$) & \multicolumn{4}{|c|}{2:40}\\
\hline  
$t_2$  & \multicolumn{4}{|c|}{$LB\over{Nv}$ = ${30 \times 40}\over{4\times v}$= $300 \over v$}\\
\hline
$T$  & 2:42+$t_2$ & 2:44+$t_2$ & 2:40+$t_2$ & 2:36+$t_2$\\
\hline                             
\end{tabular}
\end{center}
\end{table}

\begin{table}[ht]
\captionof{table}{Total number of robot = 6}   
\begin{center}                  
\small\addtolength{\tabcolsep}{-5pt}        
\begin{tabular}{|c|c|c|c|c|}             
\hline                       
Robot & Instance & Instance & Instance & Instance \\  [0.5ex]   
      & 1 & 2 & 3 & 4 \\  [0.5ex]    
      & (X,Y,O) & (X,Y,O) & (X,Y,O) & (X,Y,O) \\
\hline                              
1 &(3,-11,P) &(-11,-8,N) &(-10,-11,P) &(-7,7,P)\\               
\hline 
2 &(9,-6,P) &(17,0,N) &(15,1,N) &(3,7,N)\\ 
\hline 
3 &(14,9,P)&(14,9,N)&(10,10,N)&(8,-2,N)\\ 
\hline 
4 &(-3,-4,P)&(10,-6,N)&(-1,-8,P)&(-8,-6,P)\\
\hline 
5 &(-14,2,P)&(-14,4,N)&(-16,11,P)&(4,-6,P)\\ 
\hline 
6 &(-4,-8,P)&(7,6,N)&(-5,3,N)&(-5,2,P)\\
\hline
$t_1$ (min:sec) & 2:37 & 2:32 & 2:36 & 2:38\\          
\hline 
Avg($t_1$) & \multicolumn{4}{|c|}{2:35}\\
\hline 
$t_2$  & \multicolumn{4}{|c|}{$LB\over{Nv}$ = ${30 \times 40}\over{6\times v}$= $200 \over v$}\\
\hline 
$T$  & 2:37+$t_2$ & 2:32+$t_2$ & 2:36+$t_2$ & 2:38+$t_2$\\ 
\hline                              
\end{tabular}
\end{center}
\label{table:2}          
\end{table}

The time required by the robots to reach their destination are shown in tables \ref{table:1}, \ref{table:2} and \ref{table:3} where the 
number of robots is $4$, $6$ and $8$ respectively. In each table, four different initial instances are shown. In these instances, 
the set of robots are having different initial positions and orientations. To show the special cases, in the first two instances 
we have taken all the robots to have same orientation, either all positive or all negative. The last two instances represent 
general cases, that is, the robots are having both types of orientation. Moreover, in the fourth instance we have taken more 
than one robots on the same horizontal line. In all these cases robots reached their destination without any collision. 

The time taken to complete {\it Phase-I} is shown in the tables as $t_1$ and that to complete {\it Phase-II} is calculated and shown as $t_2$.
The results gathered from the experiments signifies that the overall job will be completed within finite amount of time. It is obvious that
as $N$ increases $t_2$ decreases. On the other hand, though it is expected that as $N$ increases $t_1$ would also decreases, instance $4$ of table \ref{table:1} 
and instance $4$ of table \ref{table:2} depicts just the reverse scenario. It is because of the fact that completion time of {\it Phase-I} is highly dependent on the 
initial distribution of the robots rather than the number of robots if they do not differ much.

\begin{table}[ht]
\captionof{table}{Total number of robot = 8}   
\begin{center}                  
\small\addtolength{\tabcolsep}{-5pt}                         
\begin{tabular}{|c|c|c|c|c|}            
\hline                       
Robot & Instance & Instance & Instance & Instance \\  [0.5ex]   
      & 1 & 2 & 3 & 4 \\  [0.5ex]    
      & (X,Y,O) & (X,Y,O) & (X,Y,O) & (X,Y,O) \\
\hline                              
1 &(0,12,P) &(0,7,N) &(-11,10,P) &(-10,9,N)\\               
\hline 
2 &(15,4,P) &(10,10,N) &(17,10,P) &(15,11,P)\\ 
\hline 
3 &(-11,8,P)&(-8,12,N)&(-13,0,P)&(-13,0,P)\\ 
\hline 
4 &(-16,-3,P)&(-16,0,N)&(-9,-8,P)&(-9,-10,N)\\
\hline 
5 &(-13,-8,P)&(-14,-9,N)&(0,-13,P)&(0,-13,P)\\ 
\hline 
6 &(6,4,P)&(15,4,N)&(-2,8,P)&(-2,11,P)\\
\hline 
7 &(7,-2,P)&(-2,-7,N)&(12,-3,N)&(12,0,N)\\ 
\hline 
8 &(11,-9,P)&(11,-10,N)&(12,-11,N)&(12,-11,N)\\
\hline
$t_1$ (min:sec) & 2:30 & 2:34 & 2:36 & 2:32\\          
\hline 
Avg($t_1$) & \multicolumn{4}{|c|}{2:33}\\
\hline 
$t_2$  & \multicolumn{4}{|c|}{$LB\over{Nv}$ = ${30 \times 40}\over{8\times v}$= $150 \over v$}\\
\hline 
$T$  & 2:30+$t_2$ & 2:34+$t_2$ & 2:36+$t_2$ & 2:32+$t_2$\\ 
\hline                              
\end{tabular}
\end{center}
\label{table:3}          
\end{table}
\section{CONCLUSION}\label{Conclusion}

This paper presents a completely distributed \emph{Painting} algorithm to paint a priori known rectangular area by $N$ no of simple, identical, autonomous, memoryless, 
mobile robots, each having their own co-ordinate systems. The robots are deployed randomly inside the rectangular area. 

This algorithm is based on standard CORDA model and {\it asynchronous} timing model. There is neither any central authority nor any external control over the robots.  
There is no communication among the robots. The algorithm guarantees complete coverage of the region without any repeated coverage and collision.\\
The same algorithm can be used to paint any other polygonal region, provided the region is convex. In that case, all the cells may have 
the same height but their area will be different. There are several scope of future research directions for this problem. Some are as follows:\\
\begin{itemize}
 \item \textbf{Environment:} The area is free of obstacles. The size and shape of the area may vary. They may be convex or concave. 
The area may or may not contain obstacles. Moreover, the shape and size of the obstacles may vary.
 \item \textbf{Visibility:}  The robots could have limited range of visibility. They can view upto a certain distance.
 \item \textbf{Model:} We have considered {\it direction-only} and {\it asynchronous} models. Other models related to direction, orientation and timing may be used to solve similar problems.
\end{itemize}

\noindent{\includegraphics[width=1in,height=1.7in,clip,keepaspectratio]{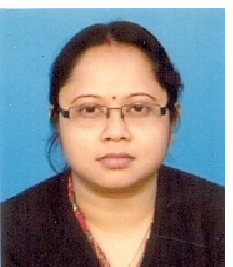}}
\begin{minipage}[b][1in][c]{1.8in}
{\centering{\bf {Deepanwita Das}} received her Bachelor of Engineering from National Institute of Technology, Durgapur in 2004. She has received her masters degree in Information Technology from Jadavpur}
\end{minipage}
University, Kolkata in 2006. She is pursuing her Ph. D. at the Department of Information Technology, National Institute of Technology, Durgapur. Currently, she is an Assistant Professor of Department of Information Technology, National Institute of Technology, 
Durgapur, West Bengal. Her research interests include swarm robotics, distributed algorithms etc.\\

\noindent{\includegraphics[width=1in,height=1.7in,clip,keepaspectratio]{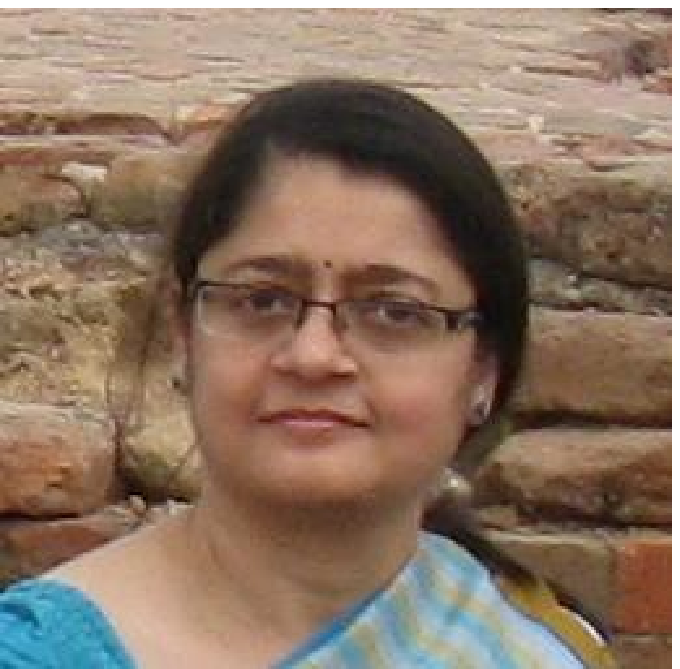}}
\begin{minipage}[b][1in][c]{1.8in}
{\centering{\bf{Srabani Mukhopadhayaya }}received B.Sc. with Hons. in Mathematics from St. Xavier’s College, Kolkata in 1987 and M.Sc. in Applied Mathematics from the University of Calcutta in 1990. She received}  
\end{minipage}
her Ph.D. in Computer Science from Indian Statistical Institute, Kolkata, in 1997. In 1998, she visited University of Florida as a Post Doctoral 
fellow. During 1999 to 2005 she was attached with Indian Statistical Institute first as a Research Associate and then as the Principal Investigator of a project under the 
Women Scientists scheme of the Department of Science and Technology, Government of India. Currently, she is an Associate Professor at Birla Institute of Technology, Mesra, 
Kolkata Campus. Her current research interests include swarm intelligence, graph and combinatorial algorithms, parallel and distributed computing, sensor networks, etc.\\\\

\end{document}